\newif\ifsingle

\newif\ifFullversion
\Fullversiontrue 

\ifsingle
\documentclass[11pt,draftclsnofoot, onecolumn]{IEEEtran}		
\else		
\documentclass[9pt,final, twocolumn]{IEEEtran}
\fi

\usepackage{times}
\usepackage{amsmath,dsfont}
\usepackage{amssymb,amsthm}
\usepackage{epsfig,verbatim}
\usepackage{subcaption}
\usepackage{setspace}
\usepackage{color}
\usepackage{cite}
\usepackage{epstopdf}
\usepackage{graphics}
\usepackage{accents}
\usepackage{acronym}
\usepackage[bookmarks,colorlinks]{hyperref}
\usepackage{booktabs}
\usepackage{mathtools}
\usepackage{enumitem}
\usepackage{multirow}
\usepackage{booktabs}
\usepackage{hyperref}
\usepackage{epstopdf} 
\usepackage{bbm}

\usepackage[ruled,linesnumbered]{algorithm2e}  
\SetKwInput{KwData}{\textbf{Initialize}} 
\usepackage{algpseudocode} 

\newcommand{\mySet}[1]{\mathcal{#1}}
	
\definecolor{NewColor}{rgb}{0,0,0} 

\ifsingle

\else

\fi 

\acrodef{adc}[ADC]{analog-to-digital convertor}
\acrodef{cs}[CS]{compressed sensing}
\acrodef{dtft}[DTFT]{discrete-time Fourier transform}
\acrodef{dnn}[NN]{neural network} 
\acrodef{csi}[CSI]{channel state information}
\acrodef{map}[MAP]{maximum a-posteriori probability}
\acrodef{snr}[SNR]{signal-to-noise ratio}
\acrodef{bs}[BS]{base station} 
\acrodef{iot}[IOT]{Interent of Things}
\acrodef{mimo}[MIMO]{multiple-input multiple-output}
\acrodef{mse}[MSE]{mean-squared error}
\acrodef{pdf}[PDF]{probability density function}
\acrodef{rv}[RV]{random variable}
\acrodef{ml}[ML]{machine learning}
\acrodef{mf}[MF]{matched filter}
\acrodef{fec}[FEC]{forward error correction}
\acrodef{rs}[RS]{Reed-Solomon}
\acrodef{lti}[LTI]{linear time-invariant}
\acrodef{wss}[WSS]{wide-sense stationary}
\acrodef{psd}[PSD]{power spectral density}
\acrodef{ser}[SER]{symbol error rate} 
\acrodef{ber}[BER]{bit error rate} 
\acrodef{sgd}[SGD]{stochastic gradient descent} 
\acrodef{isi}[ISI]{intersymbol interference}  
\acrodef{awgn}[AWGN]{additive white Gaussian noise} 
\acrodef{ut}[UT]{user terminal} 
\acrodef{mmw}[mmWave]{millimeter wave}
\acrodef{noma}[NOMA]{non-orthognal multiple access}
\acrodef{mac}[MAC]{mulitple access channel}
\acrodef{fl}[FL]{federated learning}
\acrodef{ct}[CT]{continuous-time}

\SetKwInput{KwReq}{\textbf{Input}}

\IEEEoverridecommandlockouts

\title{DRF Codes: Deep SNR-Robust Feedback Codes}

	\author{ Mahdi Boloursaz Mashhadi, Deniz G\"{u}nd\"{u}z, Alberto Perotti, and Branislav Popovic
	\thanks{M. Boloursaz Mashhadi was with the Dept. of Electrical and Electronic Eng., Imperial College London, UK. He is now with the 5GIC \& 6GIC, Institute for Communication Systems (ICS), University of Surrey, UK. D. G\"{u}nd\"{u}z is with the Dept. of Electrical and Electronic Eng., Imperial College London, UK. A. Perotti and B. Popovic are with the Radio Transmission Technology Laboratory, Huawei Technologies Sweden.}}

\vspace{-0.75cm}

\begin{document}
	
	\maketitle
	\pagestyle{empty}
	\thispagestyle{empty}
	
\begin{abstract}
We present a new deep-neural-network (DNN) based error correction code for fading channels with output feedback, called deep SNR-robust feedback (DRF) code. At the encoder, parity symbols are generated by a long short term memory (LSTM) network based on the message as well as the past forward channel outputs observed by the transmitter in a noisy fashion. The decoder  uses a bi-directional LSTM architecture along with a signal to noise ratio (SNR)-aware attention NN to decode the message. The proposed code overcomes two major shortcomings of the previously proposed DNN-based codes over channels with passive output feedback: (i) the SNR-aware attention mechanism at the decoder enables reliable application of the same trained NN over a wide range of SNR values; (ii) curriculum training with batch-size scheduling is used to speed up and stabilize training while improving the SNR-robustness of the resulting code. We show that the DRF codes significantly outperform state-of-the-art in terms of both the SNR-robustness and the error rate in additive white Gaussian noise (AWGN) channel with feedback. In fading channels with perfect phase compensation at the receiver, DRF codes learn to efficiently exploit knowledge of the instantaneous fading amplitude (which is available to the encoder through feedback) to reduce the overhead and complexity associated with channel estimation at the decoder. Finally, we show the effectiveness of DRF codes in multicast channels with feedback, where linear feedback codes are known to be strictly suboptimal.

\end{abstract}

{\textbf{\textit{Index terms---}} Communication with feedback, channel coding, LSTM, attention neural networks, curriculum training.}	


\section{Introduction}\label{sec:intro}
Most wireless communication systems incorporate some form of feedback from the receiver to the transmitter. Re-transmission mechanisms, such as hybrid automatic repeat request (HARQ), or channel state information (CSI) feedback mechanisms in time-varying channels are common forms of feedback in wireless communications. Even perfect channel output feedback does not improve the Shannon capacity of a channel \cite{Shannon1956}, but it can significantly boost the reliability at finite block-lengths \cite{SK1, S2, GN3, Polyanskiy}. Codes that make full use of feedback can potentially achieve improved performance as predicted in \cite{Polyanskiy}. However, the design of reliable codes for channels with feedback has been a long-standing and notoriously difficult open problem. Several coding schemes for channels with feedback have been proposed \cite{Horstein, SK1, Ooi, Concatenated, Ziad, Vakilinia}; however, known solutions either do not approach the performance predicted in \cite{Polyanskiy}, or introduce unaffordable complexity. These schemes are also extremely sensitive to both the precision of the numerical computations and the noise in the feedback channel \cite{S2, GN3}. It has been proven that with noisy output feedback, linear coding schemes fail to achieve any positive rate\cite{Weissman}. This is especially troubling since all practical codes are linear and linear codes are known to achieve capacity without feedback \cite{Elias}, and boost the error performance significantly in the case of noiseless feedback \cite{SK1}. For the noisy feedback case, considerable improvements have been achieved using non-linear modulo operations \cite{Shayevitz1,Shayevitz2}.

More recently, some progress has been made by applying machine learning (ML) techniques, where channel decoding is regarded as a classification task, and the encoder and decoder, implemented as deep neural network (DNN) architectures, are jointly trained in a data-driven fashion \cite{Deepcode, feedbackturbo, DEF}. In this context, the encoder/decoder pair forms an over-complete autoencoder \cite{OvercompleteAE}, where the encoder DNN adds redundancy to the latent representation of the message to cope with the channel noise, and the decoder DNN extracts features from the noisy received signal for efficient classification. In \cite{Deepcode}, the authors propose Deepcode for communication with feedback, consisting of a recurrent neural network (RNN) encoder architecture along with a two-layer bi-directional gated recurrent unit (GRU) decoder architecture, which are trained jointly on a dataset of random input/output realizations of the channel.  In \cite{feedbackturbo}, a convolutional neural network (CNN) encoder/decoder architecture with interleaving is used. In this work, the authors introduce interleaving to enable the CNN-based code to achieve a block length gain (i.e., decaying probability of error as the block length increases). In \cite{DEF}, deep extended feedback (DEF) codes are introduced, which improve the error correction capability in comparison with \cite{Deepcode} by an extended feedback mechanism that introduces longer range dependencies within the code blocks. DEF codes also enable higher spectral efficiencies by introducing higher order modulations into the encoding process. These DNN-based codes achieve lower error rates in comparison with traditional codes (i.e., Turbo, LDPC, Polar, as well as the Schalkwijk–Kailath coding scheme \cite{SK1} that exploits feedback) over the additive white Gaussian noise (AWGN) channel with output feedback at the typical code rate of $r=50/153$ and relatively short block length of $L=50$ \cite{Deepcode, feedbackturbo, DEF}. 

Despite their significant performance, DNN-based codes are very sensitive to the mismatch between the actual channel signal to noise ratio (SNR) and the SNR that the NN has been trained for, which limits their application in practical communication systems with time-varying SNR values. Although similar performance degradation is also observed with traditional channel codes when there is a mismatch between the actual channel SNR and the SNR estimation used for decoding (e.g., due to a SNR estimation error) \cite{mismatch1, mismatch2}, the impact is more critical for DNN-based codes. Since for DNN-based codes, the encoder and decoder are trained jointly, not only the decoder but also the transmitted codewords depend on the SNR. Hence, for practical deployment of the DNN-based codes on a time-varying channel, the encoder and decoder will have to train and store distinct DNNs for different SNR values, and use the appropriate one for each instantaneous value of the channel SNR, or more practically, for different ranges of SNR. This significantly increases the memory requirements and limits the practical application of DNN-based codes on realistic systems, and is a main focus of this paper.

In this paper, we propose \textbf{D}eep SNR-\textbf{R}obust \textbf{F}eedback (DRF) codes for the fading channels with noisy output feedback, which overcome the above-mentioned limitation of DNN-based channel codes. The DRF encoder transmits a message followed by a sequence of parity symbols, which are generated by a long short term memory (LSTM) architecture based on the message as well as the delayed past forward channel outputs observed by the encoder through a noisy feedback channel. The decoder uses a bi-directional LSTM architecture along with a SNR-aware attention \cite{Att1, Att2, Att3, Att4, Att5} network to decode the message. The major contributions of this paper can be summarized as follows:
\begin{itemize}

    \item We propose an attention mechanism that enables SNR-aware decoding of the DRF code, thereby considerably improving its robustness in realistic time-varying channels, where there may be a considerable mismatch between the training SNR and the instantaneous channel SNR. The attention module takes as input the forward and feedback noise variances and outputs attention coefficients that scale the features extracted at the output of the bi-directional LSTM layers at the decoder. The SNR-aware attention module significantly improves robustness of the code to a mismatch between the training and link-level SNR values in comparison with the previous works for the AWGN channel.
    
    \item We propose a training approach with SNR scheduling and batch-size adaptation. We start the training at low SNR values and with a smaller batch-size, and gradually increase the SNR and the batch-size along the training epochs according to a schedule. The proposed training approach improves the SNR-robustness of the resulting code and speeds up the training. The DRF codes and the proposed training approach not only achieve considerable SNR-robustness, but also improve the error rate over Deepcode \cite{Deepcode} roughly by an order of magnitude.
    
    \item For fading channels, in which the instantaneous SNR may be varying on each transmitted codeword (slow fading) or symbol (fast fading), we show that the proposed DRF codes learn to efficiently exploit the CSI, which is available to the encoder through feedback, such that no further improvement is possible by providing the CSI to the decoder. This is a desirable feature as it means that the complexity and overhead associated with channel estimation at the decoder can be reduced.
    
    \item For AWGN multicast channels with feedback, we show the power of DRF codes in exploiting multiple feedback signals to improve the reliability of all the receivers simultaneously. This is of significant interest as linear-feedback schemes are known to be strictly suboptimal for such channels \cite{AWGNBC1}.
    
\end{itemize}

The rest of this paper is organized as follows. In Section II, we present the feedback channel model considered in this paper. In Section III, we provide the DNN architectures for the DRF encoder and decoder. In Section IV, we present our proposed training technique. Section V presents the simulation results. Section VI extends the DRF codes to multicast channels with feedback, and Section VII concludes the paper.

\textit{\textbf{Notations:}} Throughout this paper, we denote matrices and vectors by boldface lowercase and uppercase letters, respectively. All vectors are assumed to be column vectors. The notations $(.)^T$ and $(.)^{-1}$ are used for matrix transposition and inversion, respectively. Calligraphic letters denote sets, where $|.|$ denotes cardinality of the set. Moreover, we denote the gradient by $\nabla (\cdot)$. Finally, $Pr\{.\}$ denotes probability of an event, and $\mathbb{E}{[.]}$ and $\mathrm{var}(.)$ denote the expectation and variance of random variables.

\section{System Model}\label{sec:Model}
\begin{figure}
	\centering
	\includegraphics[scale=.3]{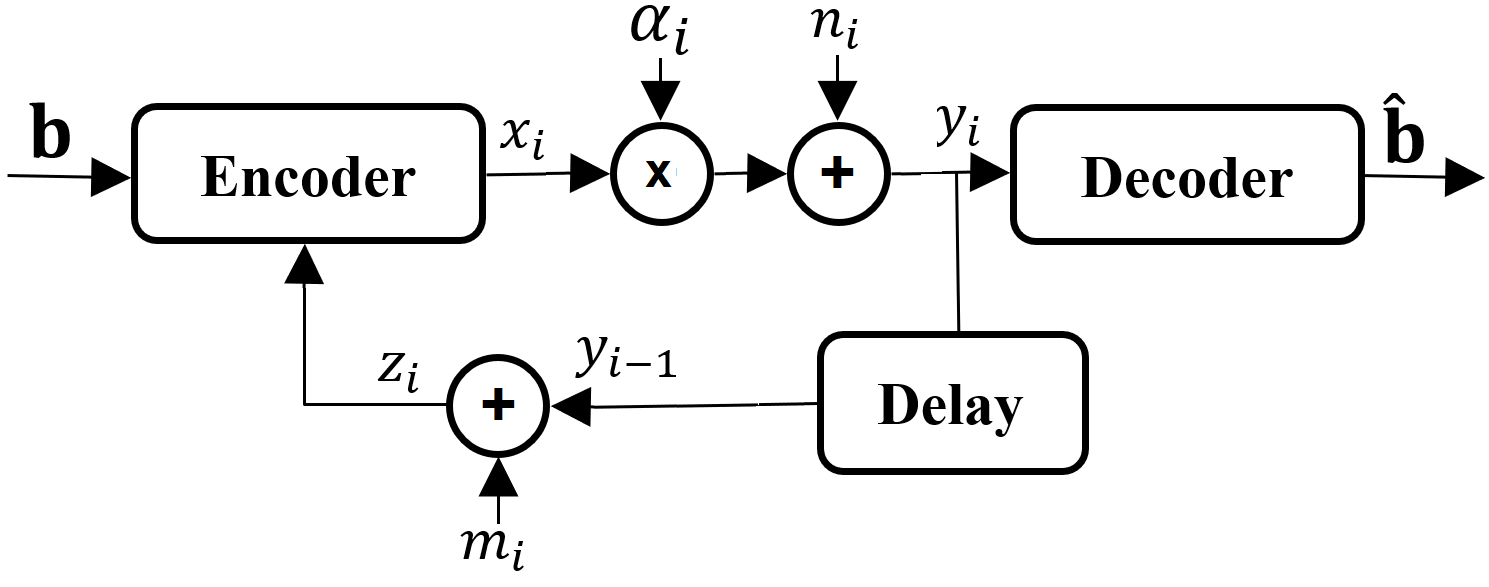} 
	\caption{Fading channel with noisy \textit{passive} output feedback.} 
	\label{fig:AWGNFeedback}
\end{figure}

Fig. \ref{fig:AWGNFeedback} illustrates the canonical fading channel passive noisy output feedback that will be the focus of this research. Perfect phase compensation at the receiver is assumed and all variables are real-valued. In this model, we have 
\begin{align}
y_{i}=\alpha_i x_{i}+n_{i},
\end{align}
where $x_{i}$ and $y_{i}$ denote the channel input and output symbols, respectively, $\alpha_i$ is the channel fading coefficient, $n_{i}$ is an independent and identically distributed i.i.d. Gaussian noise term, i.e., $n_{i}\sim\mathcal{N}(0, \sigma^2_n)$. We will assume that the channel fading coefficient comes from a prescribed distribution. We consider both slow and fast fading scenarios where the fading coefficient remains constant on each codeword for slow fading but gets i.i.d. random values on each symbol for the fast fading case. The channel output is assumed to be available at the encoder with a unit time delay via an independent AWGN feedback channel. At time $i$, the encoder has a noisy view of what was received at the decoder (in the past by one unit time)
\begin{align}
z_{i}=y_{i-1}+m_{i},
\end{align}
where $m_{i}$ is an i.i.d. Gaussian noise term, i.e., $m_{i}\sim\mathcal{N}(0, \sigma^2_m)$. We call this a \textit{passive} output feedback, as unlike in \cite{Shayevitz1,Shayevitz2}, the decoder cannot apply any coding or other type of transformation on its received signal $y_{i}$ before feeding it back to the encoder. The encoder can use the feedback symbol to sequentially and adaptively decide what to transmit as the next symbol. Therefore, channel input $x_i$ at time instant $i$ depends not only on the message $\mathbf{b}\in \{0,1\}^K$, but also on the past feedback symbols.

The encoder maps the message ${\mathbf{b}} \in \{0,1\}^K$ onto the codeword $\mathbf{x}=[x_1, \hdots, x_L]^T$, where $L$ is the block length and $K$ is the message length. The decoder maps the received codeword $\mathbf{y}=[y_1, \hdots, y_L]^T$ into the estimated information bit sequence $\hat{\mathbf{b}} \in \{0,1\}^K$., where $r = K/L$ is the rate of the code. The block error rate (BLER) is given by $\mathrm{BLER} = Pr\{\hat{\mathbf{b}} \neq \mathbf{b}\}$. Similarly, the bit error rate (BER) is given by $\mathrm{BER} = 1/K \sum_{k=1}^{K} Pr\{\hat{{b}}_k \neq {b}_k\}$, where ${b}_k$ and $\hat{{b}}_k$ denote the $k$'th bit of the transmitted and reconstructed messages, respectively. We assume an average power constraint on the channel input, i.e., $\frac{1}{L} \mathbb{E}[\|\mathbf{x}\|^2] \leq 1$, where the expectation is over the randomness in the information bits, the randomness in the noisy feedback symbols $[z_1, \hdots, z_L]^T$ and any other randomness in the encoder. We denote the forward and feedback channel SNR values by $\rho=1/\sigma^2_n$ and $\eta=1/\sigma^2_m$, respectively.

\section{Encoder/Decoder Architectures}\label{sec:Architecture}

A major limitation of the existing DNN-based code designs in \cite{Deepcode, feedbackturbo, DEF} is their dependencies on the channel SNR. That is, the encoder-decoder pairs are trained jointly for a specific SNR value. This means that, to be able to use these codes in practice, we will have to train and store a different DNN pair for different ranges of SNR values, which significantly limits their practical use in realistic channels with varying SNR. On the other hand, in conventional channel codes, the encoder depends only on the transmit power constraint, and the decoder uses the same decoding algorithm for all SNR values after converting the channel outputs into likelihood values depending on the channel SNR. Accordingly, a major goal of our paper is to implement a similar approach for DNN-based code design. This is achieved in this paper by incorporating an attention mechanism into the decoder of our proposed DRF code. This will allow us to train and store a single DNN, which can be used for all SNR values. Apart from this, we design the DRF code for fading channels with feedback, when the instantaneous channel SNR may change over time. This is different from the previous works that consider the simple AWGN channel with feedback \cite{Deepcode,feedbackturbo, DEF}. Fig. \ref{fig:Ng} depicts our proposed DRF encoder and decoder architectures for a rate $50/153$ code.

\subsection{Encoder}
\begin{figure}
\centering
\begin{subfigure}[b]{0.42\textwidth}
   \includegraphics[width=1\linewidth]{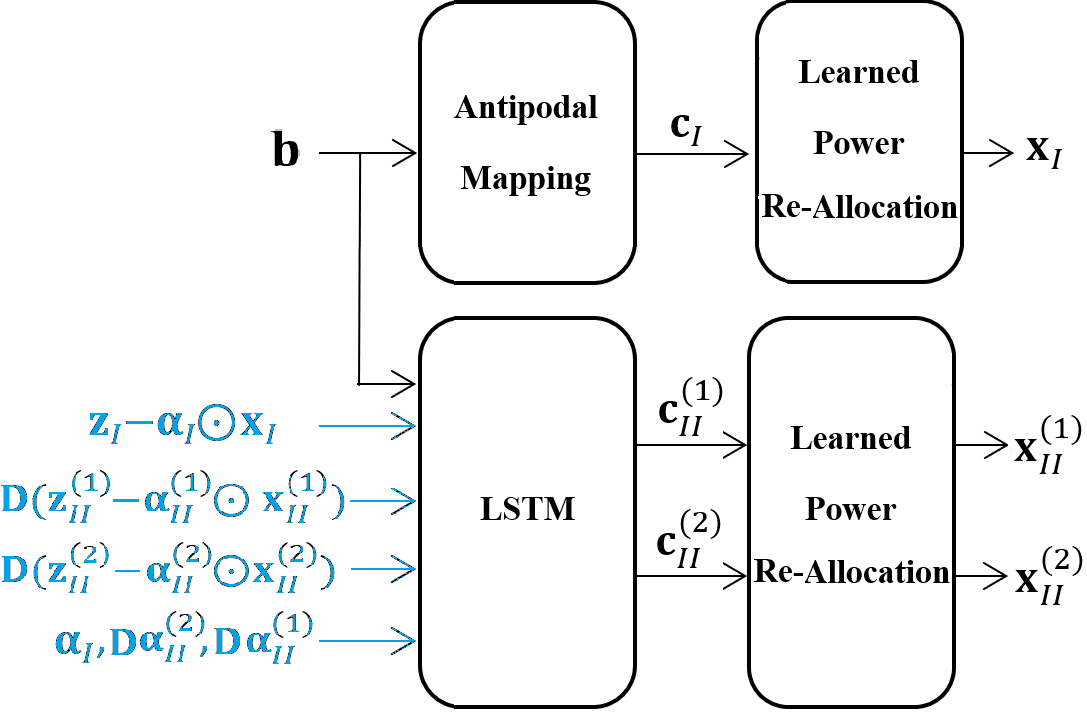}
   \caption{}
   \label{fig:Ng1} 
\end{subfigure}

\begin{subfigure}[b]{0.42\textwidth}
   \includegraphics[width=1\linewidth]{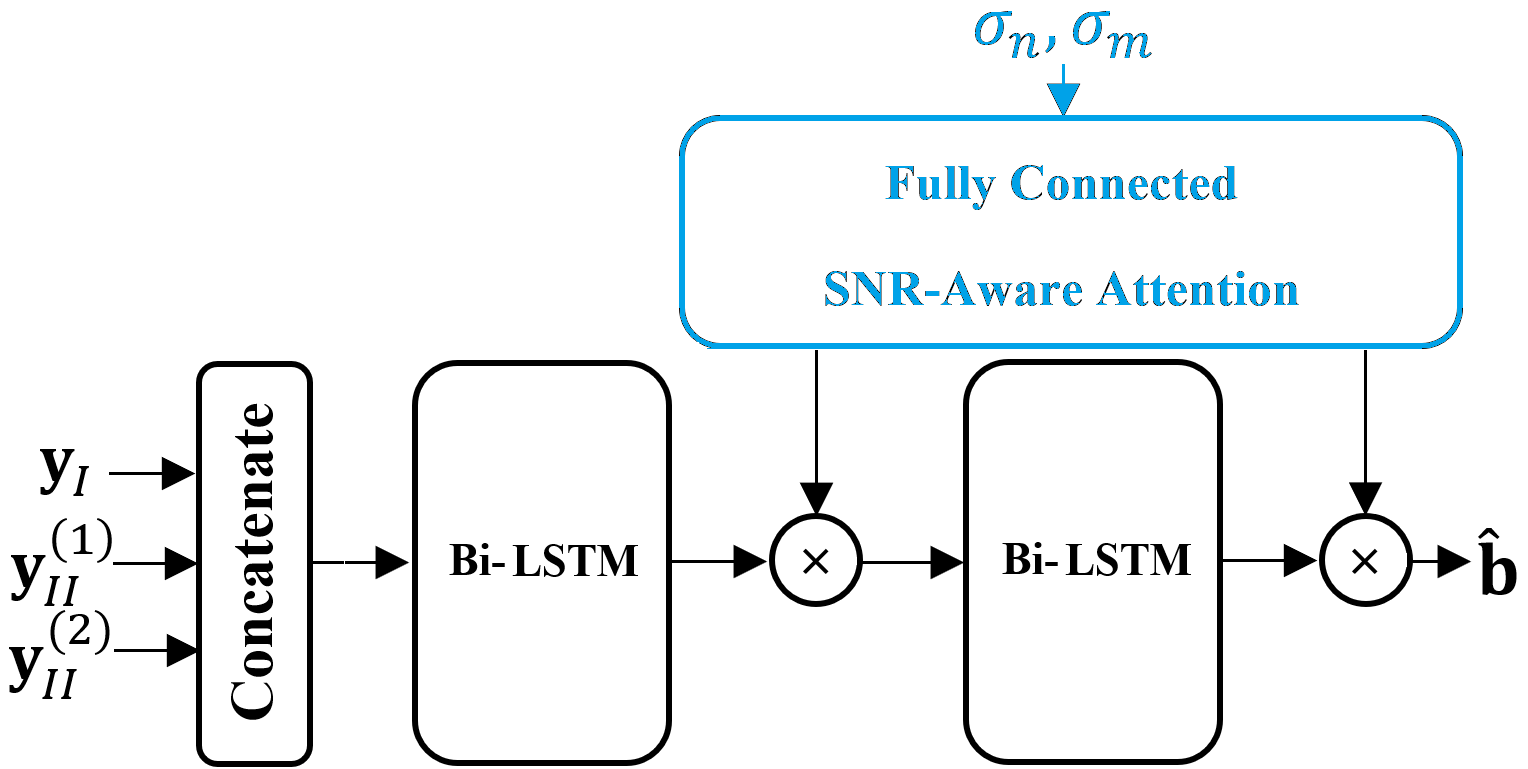}
   \caption{}
   \label{fig:Ng2}
\end{subfigure}

\caption[]{The block diagram of the proposed DRF code structure, (a) Encoder, (b) Decoder. The novel blocks of the DRF encoder and decoder architecture are shown in blue for emphasis.}
\label{fig:Ng}
\end{figure}

Fig. \ref{fig:Ng1} illustrates the encoder architecture. Encoding is a two-phase process: in phase I, the vector $\mathbf{b}=[b_1, \hdots, b_K, 0]^T$ consisting of the message bits padded by a zero is transmitted over the channel by an antipodal mapping, i.e., $\mathbf{c}_{I}=2\mathbf{b}-1$. Zero padding is applied to mitigate the increasing error rate effects on the last few bits of the block as suggested in \cite{Deepcode}. During phase II, the encoder uses a 1-layer LSTM \cite{LSTMIntro} network, including $K+1$ LSTM units to generate two sets of parity bits, i.e., $\mathbf{c}^{(1)}_{II}$ and $\mathbf{c}^{(2)}_{II}$, based on the observations of channel noise and fading in phase I and the delayed noise and fading in phase II on each of the two sets of parity symbols. We use single directional LSTM units due to the causality constraint enforced by the channel model. The LSTM activation is hyperbolic tangent, i.e., $\tanh(x)=\frac{e^{x}-e^{-x}}{e^{x}+e^{-x}}$, while the output activation function is sigmoid, i.e., $\mathrm{sigmoid}(x)=\frac{1}{1+e^{-x}}$. The resulting code block transmitted over the channel is $\mathbf{x}=[\mathbf{x}_{I}^{T}, \mathbf{x}^{(1)}_{II} {}^{T}, \mathbf{x}^{(2)}_{II} {}^{T}]^T=[x_1, \hdots,x_{3K+3}]^T$, where $\mathbf{x}_{I}=\mathcal{P}\{\mathbf{c}_{I}\}=[x_1,x_2,\hdots,x_{K+1}]^{T}$, $\mathbf{x}^{(1)}_{II}=\mathcal{P}\{\mathbf{c}^{(1)}_{II}\}=[x_{K+2},x_{K+3},\hdots,x_{2K+2}]^{T}$, and $\mathbf{x}^{(2)}_{II}=\mathcal{P}\{\mathbf{c}^{(2)}_{II}\}=[x_{2K+3},x_{2K+4},\hdots,x_{3K+3}]^{T}$. Here, $\mathcal{P}\{\cdot\}$ denotes a learned power re-allocation layer to balance the error over the whole block as suggested in \cite{Deepcode}. 

The encoder estimates the forward channel from observations of the feedback. The encoder knows the transmitted symbol $x_i$ and also observes the corresponding feedback symbol $z_{i}=\alpha_{i-1} x_{i-1} + n_{i-1}+m_{i}$ with a single delay, and therefore, it can estimate the CSI $\hat{\alpha}_{i-1}$ from its observation. This estimate of the CSI is then input to the encoder. For example, in a fast fading scenario, where the fading coefficient takes random i.i.d. realizations on each symbol, the LMMSE estimate of the channel gain can be calculated by 

\begin{equation}
    \hat{\alpha}_i=\frac{x_{i-1} \mathrm{var}(\alpha)}{|x_{i-1}|^2\mathrm{var}(\alpha)+\sigma_n^2+\sigma_m^2}{z}_{i}+\frac{\sigma_n^2+\sigma_m^2}{|x_{i-1}|^2\mathrm{var}(\alpha)+\sigma_n^2+\sigma_m^2}\mathbb{E}{[\alpha]},
    \label{eqn:LMMSEFast}
\end{equation}
where $\mathbb{E}{[\alpha]}$ and $\mathrm{var}(\alpha)$ denote the expected value and variance of the fading coefficient, respectively. In a slow fading scenario, the fading coefficient is fixed over the whole codeword, i.e., for the considered  rate $50/153$ code with a single bit zero padding we have $\alpha_1= \cdots =\alpha_{3K+3}=\alpha$. The fading coefficient $\alpha$ takes random i.i.d. realizations over different codewords, and the transmitter uses the causal vectors $\mathbf{{\mathbf{z}}}_i=[{z}_1, \hdots, {z}_{i}]^T$ and $\mathbf{x}_i=[x_1, \hdots, x_{i}]^T$, to calculate the LMMSE channel estimate as 
\begin{align}\label{eqn:LMMSESlow}
\hat{\alpha}&=\mathrm{var}(\alpha)\mathbf{x}_{i-1}^T(\mathrm{var}(\alpha)\mathbf{x}_{i-1}\mathbf{x}_{i-1}^T+\sigma_n^2I+\sigma_m^2I)^{-1}{\mathbf{z}_i}\\\nonumber
            &+\mathbb{E}{[\alpha]}(1-\mathrm{var}(\alpha)\mathbf{x}_{i-1}^T(\mathrm{var}(\alpha)\mathbf{x}_{i-1}\mathbf{x}_{i-1}^T+\sigma_n^2I+\sigma_m^2I)^{-1}\mathbf{x}_{i-1}),
\end{align}
in which $I$ is the identity matrix. In (\ref{eqn:LMMSEFast}), (\ref{eqn:LMMSESlow}), knowledge of $\mathbb{E}{[\alpha]}$ and $\mathrm{var}(\alpha)$ at the transmitter is assumed. 

The causal CSI available at the encoder is fed into the LSTM units to cope with channel uncertainty due to fading. To this end, we concatenate the vector of instantaneous channel fading coefficients in phase I, i.e. $\mathbf{\alpha}_{I}=[\alpha_{1},\hdots,\alpha_{K+1}]^T$, and the causal fading coefficient in phase II i.e. $\mathbf{D}\mathbf{\alpha}^{(1)}_{II}=[0,\alpha_{K+2},\hdots,\alpha_{2K+1}]^T$, and $\mathbf{D}\mathbf{\alpha}^{(2)}_{II}=[0,\alpha_{2K+3},\hdots,\alpha_{3K+2}]^T$, and feed into the LSTM units at the encoder ($\mathbf{D}$ denotes a single delay, see Fig. \ref{fig:Ng1}). We also provide estimates of the noise in the forward and feedback channels to the encoder, i.e. $\mathbf{z}_{I}-\alpha_{I} \odot \mathbf{x}_{I}$, $\mathbf{D}(\mathbf{z}^{(1)}_{II}-\alpha^{(1)}_{II} \odot \mathbf{x}^{(1)}_{II})$, and $\mathbf{D}(\mathbf{z}^{(2)}_{II}-\alpha^{(2)}_{II} \odot \mathbf{x}^{(2)}_{II})$, where $\odot$ denotes element-wise multiplication. For the AWGN case where $\mathbf{\alpha}_{I}=\mathbf{\alpha}^{(1)}_{II}=\mathbf{\alpha}^{(2)}_{II}=1$, the corresponding inputs are omitted to avoid unnecessary complexity.

\subsection{Decoder}
Fig. \ref{fig:Ng2} illustrates the DRF decoder consisting of a two-layer LSTM architecture (each including $K+1$ LSTM units) and a  SNR-aware fully connected attention network used for feature scaling. At the decoder, we use bi-directional LSTM layers to exploit long range forward and backward dependencies in the received code block. The phase I and II received signals are concatenated at the decoder and fed to the bi-directional LSTM layers. Each LSTM layer is followed by batch normalization. Similarly to the encoder, the LSTM activation is hyperbolic tangent while the output activation is sigmoid. The bi-directional LSTM layers extract features from the noisy received signals, which are then used for efficient decoding.

Note that we use LSTM layers at both the encoder and the decoder, which, according to our observations, considerably reduce the error rate in comparison with simple RNN and Gated Recurrent Unit (GRU) layers used in \cite{Deepcode}. This is because LSTM layers can better learn long-range dependencies by avoiding the gradient vanishing problem in training long RNN layers \cite{RNNdifficulty, RNNshort}. The LSTM architecture includes a set of gates that control when longer range information enters the memory, when it is output, and when it is forgotten \cite{LSTMIntro}. This property is very favourable for channel encoding and decoding as generating redundancies based on long range dependencies is essential to achieve a blocklengh gain.  

\begin{table}[t]
\centering
\caption{Model architecture for the SNR-aware attention module at the decoder.}
\begin{tabular}{|c|c|}
\hline
Layer           & Output Dim. \\ \hline \hline
Input           & $2$           \\ \hline
Fully connected + sigmoid & $4K^2$         \\ \hline
Fully connected + sigmoid & $2K^2$          \\ \hline
\end{tabular}
\label{tab:AWGNAttention}
\end{table}

\subsection{SNR-Aware Attention}

A major novelty in our decoder architecture is the SNR-aware attention module. An  attention  mechanism is a  vector  of  importance  weights to  measure  the correlations  between  a vector  of  inputs and the  target  to  be predicted. Attention weights are calculated as a parameterized attention function with learnable parameters \cite{Att1, Att2, Att3, Att4, Att5}.

We use a two-layer fully connected (FC) attention at the DRF decoder as outlined in Table \ref{tab:AWGNAttention}. The idea is to let the attention layers learn how much each bi-LSTM output should be weighted according to the SNR. Also, by means of the attention module, we explicitly provide the noise standard deviation to the decoder, which enables learning codes that are capable of adaptation to the channel SNR, which in turn allows to use the same trained encoder/decoder weights over a wide range of channel SNR values. Here, the standard deviations of the forward and feedback channel noise are obtained through link-level estimation. The number of attention weights determines the number of neurons at the last FC layer in Table \ref{tab:AWGNAttention} and equals $2HK$, where $H$ is the length of the LSTM hidden state (i.e., $H=K$ here) and is multiplied by $2$ because the LSTM layer is bi-directional. The total number of FC attention layers and the number of neurons in each intermediate layer are hyperparameters optimized numerically for the best performance.

\section{Training DRF Codes}\label{sec:training}

We denote the $i$'th training sample by $\mathbf{S}_i=\{\mathbf{b}_i, \mathbf{\alpha}_i, \mathbf{n}_i, \mathbf{m}_i\}$, which consists of a random realization of a message, i.e., $\mathbf{b}_i$, the corresponding realization of the channel fading coefficient $\mathbf{\alpha}_i$, and the forward and feedback noise realizations, $\mathbf{n}_i$ and $\mathbf{m}_i$, respectively. We denote the encoder and decoder functions by $f(\cdot;\mathbf{\theta})$ and $g(\cdot;\mathbf{\psi})$, where $\mathbf{\theta}$ and $\mathbf{\psi}$ are the trainable encoder and decoder parameters. We have, $\mathbf{\hat{b}}_i=g(\mathbf{\alpha}_i f(\mathbf{S}_i;\mathbf{\theta})+ \mathbf{n}_i;\mathbf{\psi})$. To train the model, we minimize 
\begin{equation}
    \mySet{L}(\mathbf{\theta}, \mathbf{\psi}, \mySet{B}) = -\frac{1}{|\mySet{B}|}\sum_{\mathbf{S}_i \in \mySet{B}}l(\mathbf{\hat{b}}_i,\mathbf{b}_i;\mathbf{\theta},\mathbf{\psi}),
    \label{eqn:NetLoss}
\end{equation}
where $\mySet{B}$ is a batch of samples, $l(\mathbf{\hat{b}}_i,\mathbf{b}_i;\mathbf{\theta},\mathbf{\psi})$ is the binary cross entropy loss given by
\begin{equation}
    l(\mathbf{\hat{b}}_i,\mathbf{b}_i;\mathbf{\theta},\mathbf{\psi})=\sum_{k=1}^{K}[\mathbf{b}_i]_k  \log_2(1-[\mathbf{\hat{b}}_i]_k)+(1-[\mathbf{b}_i]_k)  \log_2([\mathbf{\hat{b}}_i]_k),
    \label{eqn:BCELoss}
\end{equation}
and $[\mathbf{b}_i]_k$ and $[\mathbf{\hat{b}}_i]_k$ denote the $k$th bit of the message and its estimate.  

To train the model, we use a variant of the stochastic gradient descent (SGD), for which the vector of all trainable parameters $\mathbf{\phi}^T=[\mathbf{\theta}^T, \mathbf{\psi}^T]$ is updated by iterations of the form 
\begin{equation}
    \label{eqn:SGD}
    \mathbf{\phi}^{(t)} = \mathbf{\phi}^{(t-1)} - \mu_t\nabla_{\mathbf{\phi}} \mySet{L}(\mathbf{\phi}^{(t-1)}, \mySet{B}^{(t)}),
\end{equation}
where $t$ is the iteration index, $\mu_t > 0$ is the learning rate, and $\mySet{B}^{(t)}$ is a random batch from the dataset.




To ensure that the model is trained with many random realizations of the data and noise, we generate and use a new random set of samples in each epoch. We denote the dataset used in the $u$'th training epoch by $\mySet{D}^{u}=\{\mathbf{S}_i\}_{i=1}^{|\mySet{D}^{u}|}$, where $|\mySet{D}^{u}|=\zeta |\mySet{B}^{u}|$, $\zeta$ is a constant and $|\mySet{B}^{u}|$ is the batch-size for the $u$'th epoch. Training DNNs with SGD, or its variants, requires careful choice of the training parameters (e.g., learning rate, batch-size, etc.). For the specific task of training an efficient SNR-robust channel encoder and decoder, the SNR used to generate the training samples $\mathbf{S}_i$ also becomes a crucial parameter. In the following, we present our proposed training approach with SNR and batch-size scheduling which enables faster training of the DRF codes while resulting in a reliable and SNR-robust encoder/decoder pair. 

\begin{algorithm} [t] 
		\caption{Training of DRF codes with batch-size adaptation and SNR scheduling}
		\label{alg:tr}
		\KwReq{$U$, $\rho_1 \leq \rho_2 \leq \cdots \leq \rho_U$, $|\mySet{B}^{1}|$, $B_{max}$, $\zeta$, $\lambda$, $\kappa$}
		\KwData{$\mySet{L}_0=\infty$}
		\For{epoch $u = 1,2,\ldots, U$ }{
			{Randomly generate training dataset $\mySet{D}^{u}$ consisting of \\$\zeta |\mySet{B}^{u}|$ samples with forward SNR $\rho_u$\ \\ 
			Perform one epoch of training using SGD as in \eqref{eqn:SGD}} and record final loss $\mySet{L}_u$\ \\
			\eIf{$(\mySet{L}_u\ge \lambda \mySet{L}_{u-1}) \& (|\mySet{B}^{u}| < B_{max})$ }
			{Update batch-size $|\mySet{B}^{u+1}|=\kappa |\mySet{B}^{u}|$\ }{$|\mySet{B}^{u+1}|= |\mySet{B}^{u}|$\ }
	} 
	\KwOut{Trained encoder/decoder parameters $\mathbf{\theta}, \mathbf{\psi}$}
\end{algorithm}

\subsection{Batch-size Adaptation} In training machine learning models, a static batch-size held constant throughout the training process forces the user to resolve a tradeoff. On one hand, small batch sizes are desirable since they tend to achieve faster convergence. On the other hand, large batch sizes offer more data-parallelism, which in turn improves computational efficiency and scalability \cite{keskar, Goyal}. However, for the specific channel encoder/decoder training task a significantly larger batch size is necessary not only due to the data-parallelism benefits, but also because after a few training steps, the error rate and consequently the binary cross entropy loss (\ref{eqn:NetLoss}) becomes very small, typically $ 10^{-4} \sim 10^{-7}$ for the range of SNR values considered here. Hence, to get a statistically accurate estimate of such a small loss value, and consequently, an accurate estimate of the gradient update in (\ref{eqn:SGD}), the batch-size must be very large (typically $\sim 10000$ samples here). 

\begin{figure*}
	\centering
	\includegraphics[scale=.35]{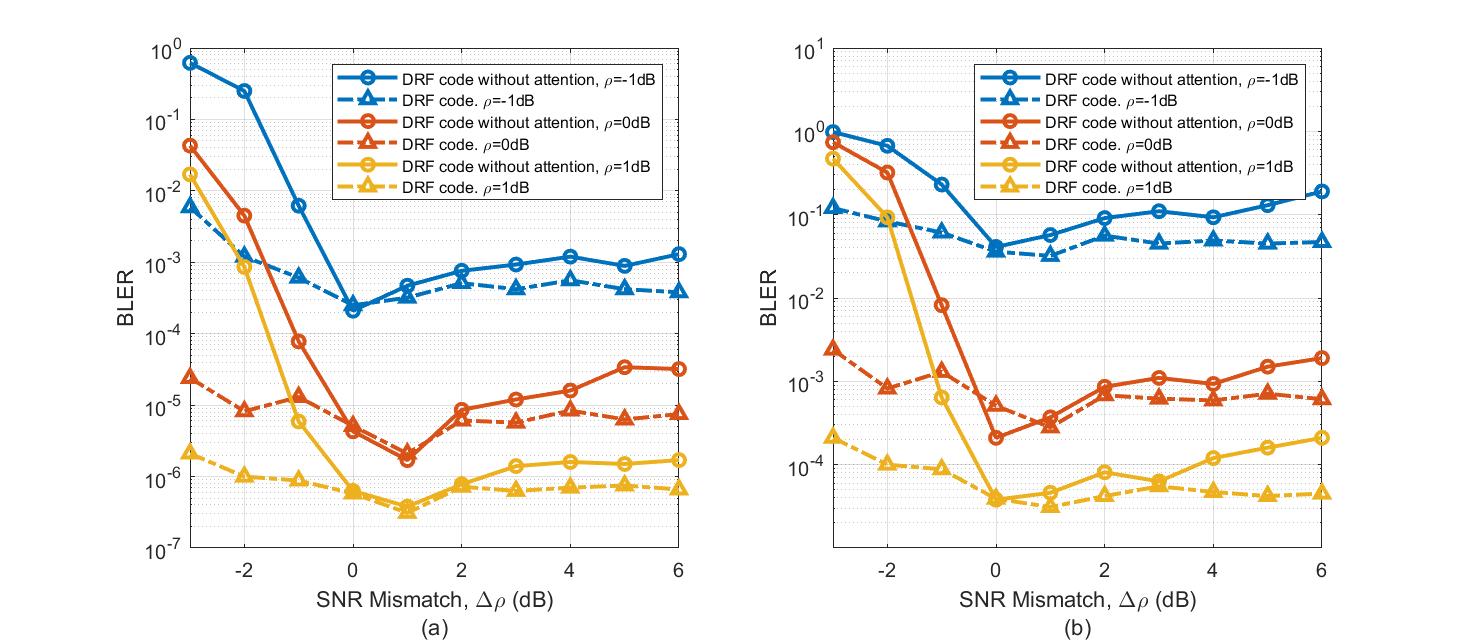} 
	\caption{Comparison between the DRF code and LSTM-based Deepcode in terms of BLER as a function of SNR mismatch $\Delta\rho$, (a) Noiseless feedback ($\eta=\infty$), (b) Noisy feedback ($\eta=20$dB).} 
	\label{fig:SNRRangePerf}
\end{figure*}

If the batch-size is small, the performance saturates after several epochs meaning that  the optimizer enters a state when it keeps iterating with inaccurate gradient estimates leading to random fluctuations in the loss value with no actual improvement. This is due to the fact that a small batch-size cannot provide an accurate estimate of the gradient, and hence, training is stuck oscillating around a minimum value, but is not able to further approach it. In this situation, as the batch size is not sufficiently large to estimate the gradient accurately, an unfortunate random realization of any batch can lead to a destructive update of the model parameters causing sudden jumps in the loss function and the BLER along the training. It was observed in simulations that such destructive updates hamper convergence, and at times, can totally destroy the code and result in divergence.

To avoid these destructive updates, we here propose an adaptive batch size scheme tailored for training a DNN-based channel encoder and decoder pair. In this scheme, we train the model starting from a small batch-size $|\mySet{B}^{1}|$, and multiply the batch size by a factor of $\kappa$ whenever the cross entropy loss does not decrease by a factor of $\lambda$ in two consecutive epochs, until we reach a maximum batch-size value $B_{max}$. The maximum batch-size is constrained by the memory resources available to our training platform. We hence train with a sequence of batch-sizes, $|\mySet{B}^{1}| \leq |\mySet{B}^{2}| \leq \cdots \leq |\mySet{B}^{U}| \leq B_{max}$, where $U$ is the total number of epochs. Starting from a smaller batch size enables a faster convergence during initial epochs. We increase the batch size whenever trapped around a minimum due to insufficiency of the batch size to achieve an accurate estimate of the gradient. This way, destructive updates become less likely as we avoid iterating with inaccurate gradient estimates. The proposed batch-size adaptation stabilizes and speeds up the training process by avoiding destructive updates due to batch-size insufficiency.

\subsection{SNR Scheduling} When training the channel encoder/decoder pair for a range of SNR values, if low and high SNR samples are presented to the decoder together during training, the trained NN tends to be biased towards the lower SNR. This is because the error probability for higher SNR values can be orders of magnitude smaller than the lower ones. Hence, the contribution of the high SNR samples in the batch to the binary cross entropy loss (\ref{eqn:NetLoss}) becomes negligible. In this case, the low SNR samples will decide the loss value and consequently the gradient updates (\ref{eqn:SGD}) causing the channel code to be biased towards lower SNR values. On the other hand, training is easier for lower SNR values, in the sense that, for higher SNR values, destructive updates become more frequent causing the training to become less stable.

To train a channel encoder and decoder pair suitable for a wide SNR range, we here propose a scheduled-SNR training approach. This is motivated by the idea of curriculum training  \cite{Curriculum1, Curriculum2}, which suggests using a ``curriculum'' in presenting training samples to the DNN based on their ``difficulty''. Curriculum training improves both the speed of convergence of the training process, and the quality of the local minima obtained in the case of non-convex optimization criteria \cite{Curriculum1, Curriculum2}. 

Assume the goal is to efficiently train a channel encoder/decoder pair that works sufficiently well for all forward channel SNR values $\rho \in [\rho_{min},\rho_{max}]$. We start training with lower SNR samples and increase the SNR along the epochs using a  SNR schedule of $\rho_{min}=\rho_1 \leq \rho_2 \leq \cdots \leq \rho_U=\rho_{max}$. We observed that SNR scheduling combined with batch-size adaptation not only stabilizes and speeds up the training, but also improves the SNR-robustness when training an encoder/decoder pair for a wider SNR range. Algorithm \ref{alg:tr} summarizes our training approach for DRF codes. The hyperparameters $U$, $|\mySet{B}^{1}|$, $B_{max}$, $\zeta$, $\lambda$, and $\kappa$ are chosen by numerical evaluations for the best performance.


\section{Numerical Evaluations}\label{sec:results}
In this section, we evaluate the performance of the proposed DRF codes and provide comparisons with previous works. In all simulations, we use $10^9$ random samples to achieve a reliable estimate of the error rate. Each sample includes a random realization of the message $\mathbf{b}$, and the corresponding random realizations of forward and feedback channels. In all the simulations, we set $K=50, L=153$, and the NN optimizer is Adam \cite{kingma2014adam}. The values of the hyperparameters are: $U=15$, $|\mySet{B}^{1}|=1000$, $B_{max}=16000$, $\zeta=100$, $\lambda=2$, $\kappa=2$.

\begin{table*}
\centering
\caption{SNR robustness for the proposed DRF codes and comparison with Deepcode, noiseless feedback ($\eta=\infty$).}
\resizebox{16cm}{!}{
\begin{tabular}{|c|c|c|c|c|} 
\hline
Test SNR & $-1$dB & $0$dB & $1$dB & $2$dB \\ 
\hline
\hline
DRF code without attention (Separate Trained DNNs)  & $2.1 \times 10^{-4}$ & $\mathbf{1.7 \times 10^{-6}}$ & $3.8 \times 10^{-7}$ & $5.6 \times 10^{-8}$\\
\hline
DRF code without attention (Trained over $[-1,2]$dB) & $2.8 \times 10^{-4}$ & $7.9 \times 10^{-6}$ & $1.6 \times 10^{-6}$ & $4.8 \times 10^{-7}$ \\
\hline
DRF code with attention (Trained over $[-1,2]$dB) & $\mathbf{1.8 \times 10^{-4}}$ & $5.3 \times 10^{-6}$ & $9.7 \times 10^{-7}$ & $1.2 \times 10^{-7}$\\
\hline
DRF code (SNR Scheduling) & $2.0 \times 10^{-4}$ & $2.4 \times 10^{-6}$ & $\mathbf{2.9 \times 10^{-7}}$ & $\mathbf{5.1 \times 10^{-8}}$ \\
\hline
\end{tabular}}
\label{tab:SNRRangePerf}
\vspace{-0.2cm}
\end{table*}


\subsection{AWGN Channel}
In this subsection, we consider the AWGN case, i.e. $\alpha_i=1, \forall i$. We first show the robustness of the proposed DRF codes to a mismatch between the training and the actual channel SNR values. We then provide an ablation study to separately show effectiveness of the SNR-aware attention mechanism and the proposed training approach to achieve SNR robustness. We finally provide BLER comparisons between the DRF codes and both the conventional and feedback channel codes. We show that DRF codes outperform the benchmark low density parity check (LDPC) codes adopted for the fifth generation new radio (5G NR) \cite{NRLDPC}, by three orders of magnitude and the previously proposed Deepcode \cite{Deepcode} by an order of magnitude. 

\subsubsection{\textbf{SNR-Robustness}}
We first compare the BLER of the proposed DRF codes with and without the attention module, when there is a mismatch between the actual channel SNR and the SNR used for training. Here, we train the codes with batch-size adaption but for a specific SNR value (i.e., without SNR scheduling). The SNR mismatch is defined as $\Delta \rho = \rho - \hat{\rho}$, where $\rho$ is the actual channel SNR and $\hat{\rho}$ is the SNR used for training. 

The results are depicted in Fig. \ref{fig:SNRRangePerf}, where we plot the BLER versus $\Delta \rho$ for $\rho=-1, 0, 1 $ dB. The results are plotted for both noiseless $\eta=\infty$ and noisy feedback at $\eta=20 $ dB. This figure shows that without the SNR-aware attention module at the decoder, the BLER is very sensitive to the SNR mismatch. In this case, a negative SNR mismatch (i.e., training SNR is higher than the actual channel SNR), can significantly degrade the BLER by orders of magnitude. The BLER is less sensitive to a positive mismatch but still roughly an order of magnitude BLER degradation is observed if there is $\Delta \rho = +3$ dB mismatch between the training and test SNR values. This figure shows that DRF codes are significantly more robust to both positive and negative SNR mismatch due to the SNR-aware attention layers added to the decoder.               

\subsubsection{\textbf{Ablation Study}}
In this subsection, we compare the BLER of the DRF architecture with and without SNR scheduling and attention, when the goal is to train a single DNN that works sufficiently well over all the SNR values $\rho \in \{-1, 0, 1, 2\}$ dB. The idea is to show the effectiveness of the proposed SNR-aware attention at the decoder and scheduling of the training SNR, separately. To this end, we report in Table \ref{tab:SNRRangePerf}, the BLER values versus channel SNR when the feedback link is noiseless. We train all the schemes with batch-size adaptation. The results labeled as ``DRF code without attention (Separate Trained DNNs)" in Table \ref{tab:SNRRangePerf}, shows the performance when we have trained 4 different DNNs (without the attention module) at forward SNR values of $\rho \in \{-1, 0, 1, 2\}$ dB, and evaluated them on the same test SNR value. However, in a realistic time-varying channel, the instantaneous SNR varies with time (e.g., a slow channel fading scenario). In such cases, switching between separate DNNs for channel encoding/decoding is less practical. 

\begin{figure}
	\centering
	\includegraphics[scale=.4]{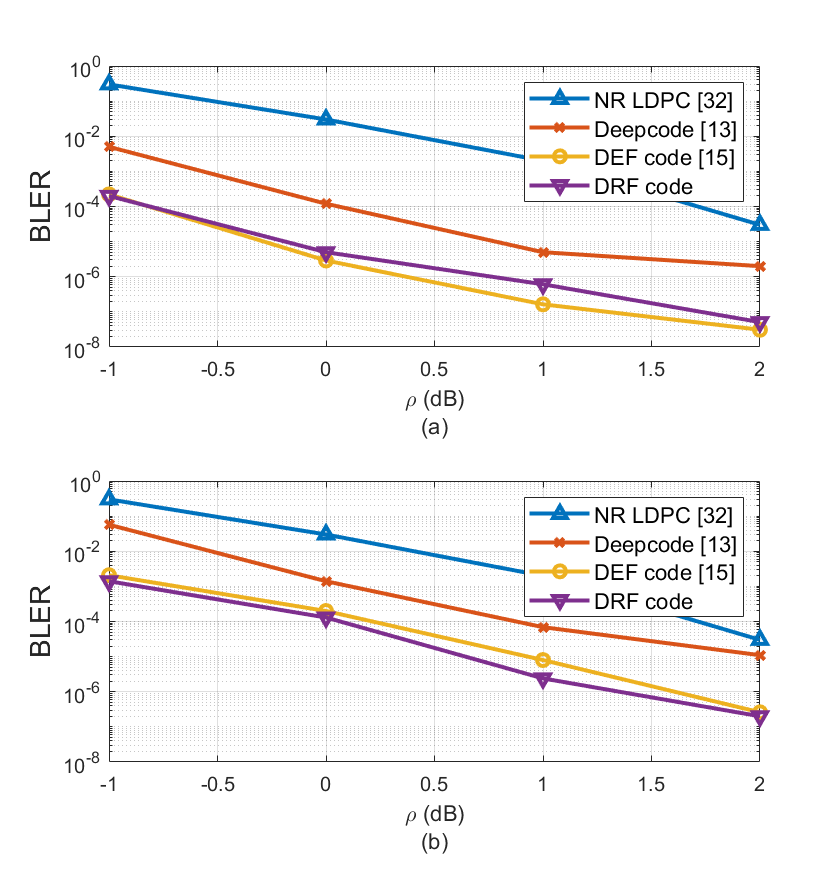} 
	\caption{Comparison between the proposed DRF codes and previous works, (a) Noiseless feedback ($\eta=\infty$), (b) Noisy feedback ($\eta=20$dB).} 
	\label{fig:AWGNRes}
\end{figure}

The results labeled as ``DRF code without attention (Trained over $[-1,2]$dB)" in Table \ref{tab:SNRRangePerf}, correspond to the scheme where we train a single DRF code architecture without attention on training samples generated with SNR values picked uniformly at random from $[-1, 2]$dB and then tested on each SNR value. As shown in Table \ref{tab:SNRRangePerf}, this approach leads to considerable performance degradation specifically at higher SNR values. This is because the trained code is considerably biased towards a better performance at low SNR values.

The results labeled as ``DRF code with attention (Trained over $[-1,2]$dB)" in Table \ref{tab:SNRRangePerf}, report the performance when a single DRF architecture (including the attention module) is trained on samples generated with SNR values picked uniformly at random from $[-1, 2]$dB and then tested on each SNR value. As the DRF decoder is aware of the SNR, it does not suffer as much performance degradation at high SNR values when it is trained over random SNR values. However, it is still slightly biased towards a better performance at low SNR values.

The results labeled as ``DRF code (SNR Scheduling)" in Table \ref{tab:SNRRangePerf}, report the performance when a single DRF code architecture (including the attention module) is trained with the proposed SNR scheduling approach. Here, instead of training with samples generated with random SNR values picked from $[-1, 2]$dB, we train for 3 epochs on samples generated with each of the SNR values in the schedule ``-1, -1, 0, 1, 2" dB, respectively, in that order (we observed in simulations that more training at SNR -1 dB improves the final performance). Comparing the last two rows of Table \ref{tab:SNRRangePerf}, we observe that curriculum training with a  SNR schedule further improves the performance, specifically for the higher SNR values. 

According to these results, the proposed DRF code architecture along with SNR scheduling achieves BLER better than or comparable with the ``Separate Trained DNNs" case while alleviating the need to train and store several DNNs for various SNR values, thereby, significantly improving practicality of the DNN-based code. Similar results are observed for the noisy feedback case with $\eta=20$ dB.


\begin{figure}
	\centering
	\includegraphics[scale=.6]{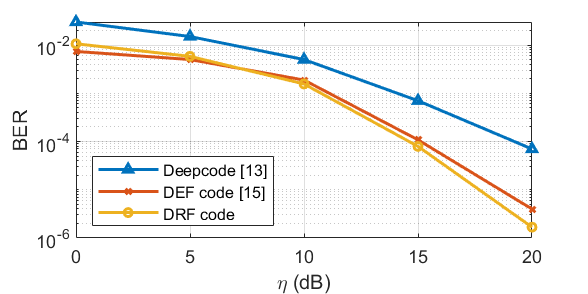} 
	\caption{BER curves versus $\eta$ for forward SNR $\rho=0$dB.} 
	\label{fig:AWGNResetta}
\end{figure}

\subsubsection{\textbf{Comparison with Previous Works}}
In this subsection, we compare the performance of DRF codes with NR LDPC \cite{NRLDPC}, Deepcode \cite{Deepcode}, and the DEF code \cite{DEF}. Fig. \ref{fig:AWGNRes} compares the BLER values achieved for each code for the forward channel SNR values in the range $\{-1,2\}$ dB when (a) the feedback is noiseless ($\eta=\infty$), and (b) the feedback SNR is $\eta=20$ dB. The blue curve reports the BLER for the RNN-based Deepcode architecture as proposed in \cite{Deepcode}. According to this figure, the proposed DRF codes reduce the BLER by almost three orders of magnitude in comparison with NR LDPC and an order of magnitude in comparison with Deepcode \cite{Deepcode}. Note that for the Deepcode and DEF code, we have trained and used a different DNN for each of the four SNR points. However, for the DRF code, we have used a single DNN for all the SNR points, which is trained using our proposed SNR scheduling approach. Hence, in comparison with the state-of-the-art DEF code, DRF code achieves SNR-robustness with no significant performance degradation.

We note that although DNN-based codes achieve huge BLER reductions for the low SNR values considered here, we do not observe a decay as fast as the traditional channel codes (e.g., LDPC) in their error rate as the SNR increases. This may be due to the fact that at higher SNR values, the error rate and consequently the binary cross entropy loss becomes too small to be accurately estimated with affordable batch-size values making the training unstable as mentioned in Section \ref{sec:training}. The error rate decay in the high SNR for the DNN-based codes will be further investigated in future research. Finally, we plot the BER versus feedback SNR curves in Fig. \ref{fig:AWGNResetta} when the forward SNR value is fixed at $\rho=0$dB. The proposed DRF code outperforms both Deepcode and DEF code. 


\subsection{Fading Channel}
In this subsection, we consider fading channels with feedback as depicted in Fig. \ref{fig:AWGNFeedback}. Depending on the wireless environment, the CSI coefficient $\alpha_i$ may follow various statistics \cite{biglieri1998fading, simon1998unified}.  In this section we adopt the Rayleigh channel assumption, which is valid for rich scattering urban environments when there exists no dominant line-of-sight (LoS) multipath component. Hence, the forward channel gain $\alpha_i$ follows the probability density function (pdf) $f(\alpha)=\frac{\alpha}{\sigma^2}e^{\frac{-\alpha^2}{2\sigma^2}}, \alpha>0$, where $\sigma$ is the scale parameter and the average power gain of the fading channel is given by $\Omega=2\sigma^2$.


\begin{figure}
	\centering
	\includegraphics[scale=.4]{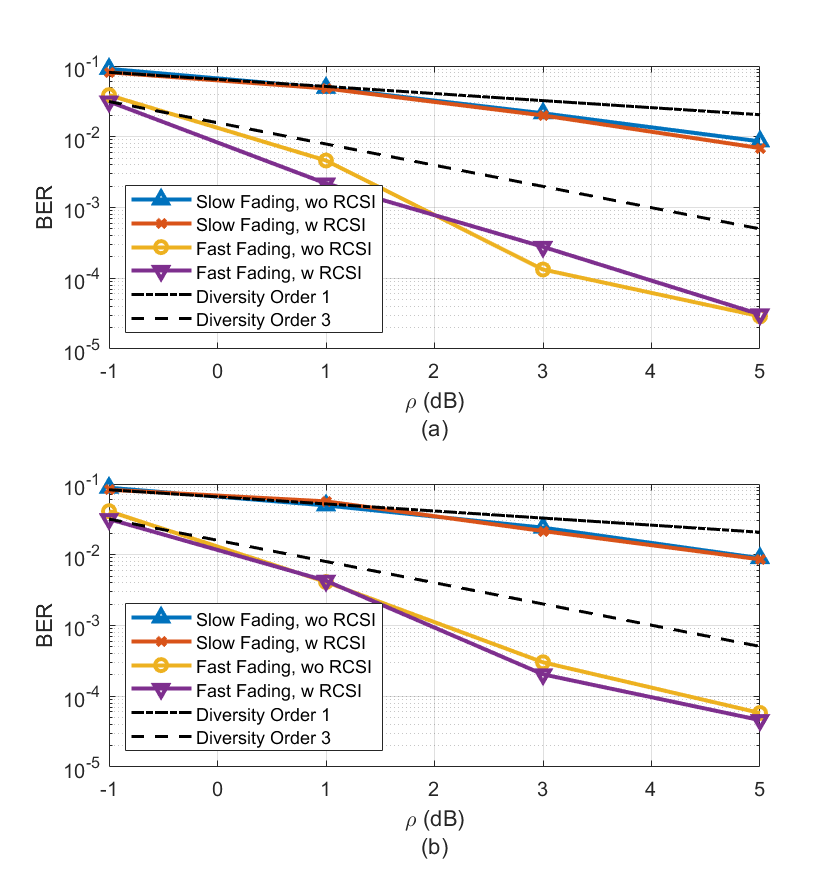} 
	\caption{The BER curves for DRF codes over Rayleigh magnitude fading channels with $\Omega=2\sigma^2=1$ as a function of the average channel SNR, $\rho$ (dB). (a) Noiseless feedback ($\eta=\infty$), (b) Noisy feedback ($\eta=20$dB).} 
	\label{fig:RCSIvsNoRCSI}
\end{figure}

We consider both slow and fast fading scenarios. In the slow fading case, the fading coefficient takes random i.i.d. Rayleigh realizations over each transmitted codeword, but remains constant throughout the transmission of the codeword. In the fast fading case, fading coefficients take random i.i.d. Raleigh realizations on each transmitted symbol. This is an extreme case as we assume consecutive channel realizations to be independent, while these are typically correlated in practice, governed by the Doppler profile of the channel.


In Fig. \ref{fig:RCSIvsNoRCSI}, we compare the resulting bit error rate (BER) curves for DRF encoding/decoding over Rayleigh magnitude fading channels when the CSI is and is not available at the receiver (i.e. with and without RCSI). In the case with RCSI, the decoder first performs linear minimum mean square error (LMMSE) channel compensation on the received symbols, i.e., $\hat{y}_i=\frac{\alpha_i}{|\alpha_i|^2+\sigma_n^2}$, and then uses $\hat{y}_i$ as input to the bi-directional LSTM units for decoding. Note that the encoder is the same as depicted in Fig. \ref{fig:Ng1} for both cases. Fig. \color{red}6a\color{black} \ exhibits the BER curves for the noiseless feedback case ($\eta=\infty$), and Fig. \color{red}6b\color{black} \   for the noisy feedback case at $\eta=20$dB. For a fair comparison, we use the exact value of the $\alpha_i$ (not an estimated version) both at the encoder and decoder. The curves show similar performance for the two cases with and without CSI at the decoder for both the slow and fast fading cases. In other words, the proposed DRF code learns to efficiently exploit the knowledge of the instantaneous CSI value $\alpha_i$ available to the encoder through feedback, such that no further improvement is achieved by providing the CSI also to the decoder. In other words, regardless of the fading scenario (i.e., slow or fast fading), providing the decoder with the knowledge of perfect instantaneous CSI does not achieve any further improvement in the error rate. This is a desirable result as it shows that using the proposed DRF codes, the complexity and overhead associated with channel estimation at the decoder can be reduced. Finally, note that the dotted curves in Fig. \ref{fig:RCSIvsNoRCSI} represent the tangent lines with slopes corresponding to diversity orders $1$ and $3$ for comparison. The DRF codes achieve considerably better diversity orders compared with the tangent lines specifically in the fast fading case.    

\begin{table*}
\centering
\caption{BLER values for the two user AWGN multicast channel with noiseless output feedback ($\eta_1=\eta_2=\infty$).}
\resizebox{16cm}{!}{
\begin{tabular}{|c|c|c|c|c|} 
\hline
SNR Pair $(\rho_1,\rho_2)$ & $(0,0)$ dB & $(0,2)$ dB & $(2,2)$ dB & $(2,0)$ dB \\ 
\hline
\hline
$\epsilon=0$  & $(3.3\times 10^{-1},3.2\times 10^{-1})$ & $(3.2\times 10^{-2},1.5\times 10^{-2})$ & $(1.6\times 10^{-4},2.2\times 10^{-4})$ & $(5.3\times 10^{-3},1.2\times 10^{-2})$\\
\hline
$\epsilon=0.9$ & $(4.7\times 10^{-3},3.8\times 10^{-3})$ & $(1.2\times 10^{-3},3.4\times 10^{-4})$ & $(9.0\times 10^{-6},2.4\times 10^{-5})$ & $(2.5\times 10^{-4},1.0\times 10^{-3})$ \\
\hline
$\epsilon=-0.9$ & $(1.3\times 10^{-1},1.3\times 10^{-1})$ & $(6.3\times 10^{-3},1.9\times 10^{-3})$ & $(4.5\times 10^{-5},4.1\times 10^{-5})$ & $(5.6\times 10^{-3},1.0\times 10^{-2})$\\
\hline
Point to point bound & $(2.0\times 10^{-4},2.0\times 10^{-4})$ & $(2.0\times 10^{-4},5.1\times 10^{-8})$ & $(5.1\times 10^{-8},5.1\times 10^{-8})$ & $(5.1\times 10^{-8},2.0\times 10^{-4})$ \\
\hline
\end{tabular}}
\label{tab:AWGNBCPerf}
\vspace{-0.2cm}
\end{table*}

\section{Multicast Channels with Feedback}
In a multicast channel with feedback the goal is to transmit a common message to multiple recievers simultaneously while exploiting seprate feedback signals from the decoders. This is another example of channels for which we do not have efficient codes with theoretical guarantees. It has been shown that linear feedback approaches that achieve the capacity and improve the error exponent in the case of a single receiver, are strictly suboptimal in the case of multicasting \cite{AWGNBC1}. Linear feedback approaches, even with perfect feedback, fail to achieve the capacity. In the extreme case where the number of receivers goes to infinity, the largest rate achieved by linear feedback schemes tends to zero \cite{AWGNBC1}. It is clear that non-linear coding schemes are necessary for this channel; however, designing such codes is meticulously difficult. In this section, we show that DRF codes can be employed for multicasting a common message to multiple receivers with a noisy output feedback from each receiver.

\begin{figure}
	\centering
	\includegraphics[scale=.33]{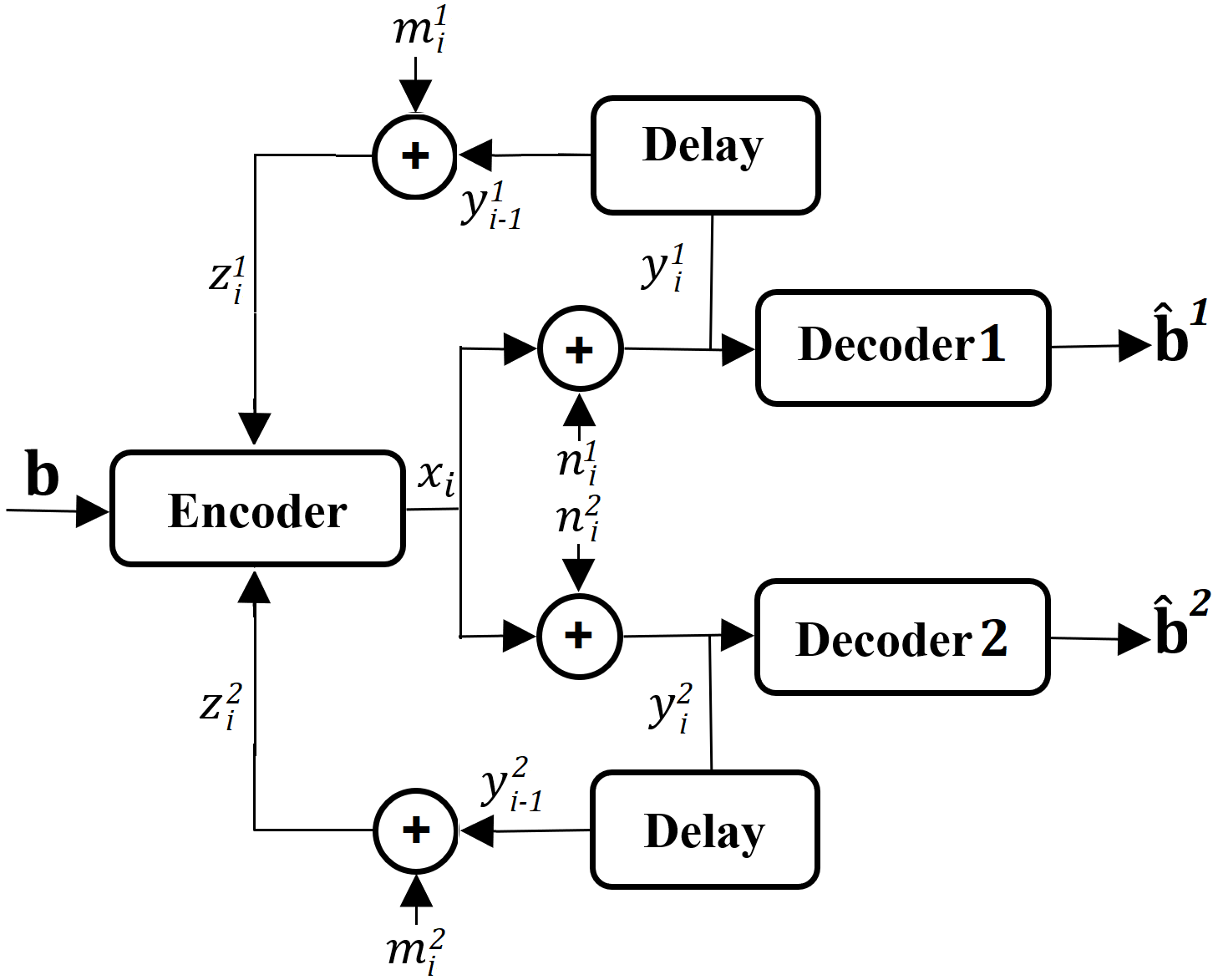} 
	\caption{The two-user AWGN multicast channel with noisy output feedback and common message.} 
	\label{fig:AWGNBC}
\end{figure}

Fig. \ref{fig:AWGNBC} shows the channel model for a multicast AWGN channel with feedback, where the encoder transmits the common message $\mathbf{b} \in \{0,1\}^K$, where $K$ is the message length, to two receivers simultaneously. In this model, $x_{i}$ denotes the channel input and $y_{i}^1$ and $y_{i}^2$ are the channel output symbols at receivers 1 and 2, respectively, where $y_{i}^1=x_{i}+n_{i}^1$ and $y_{i}^2=x_{i}+n_{i}^2$, and $n_{i}^1$ and $n_{i}^2$ are jointly Gaussian noise terms, with variances $\sigma^2_{n^1}$ and $\sigma^2_{n^2}$ and correlation coefficient $\epsilon$. The two channel outputs are assumed to be available at the encoder with a unit time delay via passive AWGN feedback channels. At time $i$, the encoder has a noisy view of what was received by both receivers: $z_{i}^1=y_{i-1}^1+m_{i}^1$ and $z_{i}^2=y_{i-1}^2+m_{i}^2$, where $m_{i}^1$ and $m_{i}^2$ are i.i.d. Gaussian noise terms, i.e., $m_{i-1}^1\sim\mathcal{N}(0, \sigma^2_{m^1})$ and $m_{i-1}^2\sim\mathcal{N}(0, \sigma^2_{m^2})$. The encoder can use the feedback symbols $z_{i}^1, z_{i}^2$ from the two receivers to sequentially and adaptively decide what to transmit as the next symbol.

The encoder maps the message ${\mathbf{b}} \in \{0,1\}^K$ onto the codeword $\mathbf{x}=[x_1, \hdots, x_L]^T$, where $L$ is the block length and $K$ is the message length. The two decoders map their received codewords $[y_{1}^1, \hdots, y_{L}^1]^T$ and $[y_{1}^2, \hdots, y_{L}^2]^T$ into the estimated information bits $\hat{\mathbf{b}}^1 \in \{0,1\}^K$ and $\hat{\mathbf{b}}^2 \in \{0,1\}^K$, where $r = K/L$ is the rate of the code. The block error probabilities for the two receivers are given by $Pr\{\hat{\mathbf{b}}^1 \neq \mathbf{b}\}$ and $Pr\{\hat{\mathbf{b}}^2 \neq \mathbf{b}\}$, respectively. As before, we impose an average power
constraint on the channel input, i.e., $\frac{1}{L} \mathbb{E}[\|\mathbf{x}\|^2] \leq 1$, where the expectation is over the randomness in the information bits, the randomness in the noisy feedback symbols and any other randomness in the encoder. We denote the forward and feedback channel SNR values for receiver $r$ by $\rho^r=1/\sigma^2_{n^r}$, and $\eta^r=1/\sigma^2_{m^r}$, $r=1,2$.

The proposed DRF codes provide powerful tools for designing efficient codes for such channels even in the case of noisy feedback. This is achieved by a slight modification of the encoder network to enable the encoder receive as input the feedback symbols from multiple receivers. The LSTM units determine the parity symbols based on the received feedback from both receivers. This is a challenging task depending on how correlated the two forward noise terms, $n_{1i}$ and $n_{2i}$, are. The decoder is a two-layer bidirectional LSTM architecture as in the point-to-point case. The loss function is the summation of the binary cross entropy losses at the two decoders. In general, a weighted sum can be considered to give priority to one of the receivers over the other. We later show through simulations the power of DRF codes in exploiting multiple feedback signals to improve the reliability at both receivers. 


In Table \ref{tab:AWGNBCPerf}, we report the BLER pairs achieved for the forward SNR pairs of $(\rho_1,\rho_2) = (0,0), (2,0), (0,2), (2,2)$ dB when the correlation coefficient between the two forward noise sequences is $\epsilon = \{0, 0.9, -0.9\}$. Here, the code rate is $r=50/153$ and the feedback from both receivers is noiseless. We also provide the BLER values for the point-to-point (single user) case for reference. The two user BLER values considerably degrade in comparison with the point-to-point case. As expected, the BLER degradation is most when the two forward channel noise sequences are independent. This is due to the fact that when generating parity symbols, the LSTM cells will have to compromise between correcting errors for the two receivers. When these errors are independent, this compromise becomes the most challenging.

\begin{figure}
	\centering
	\includegraphics[scale=.4]{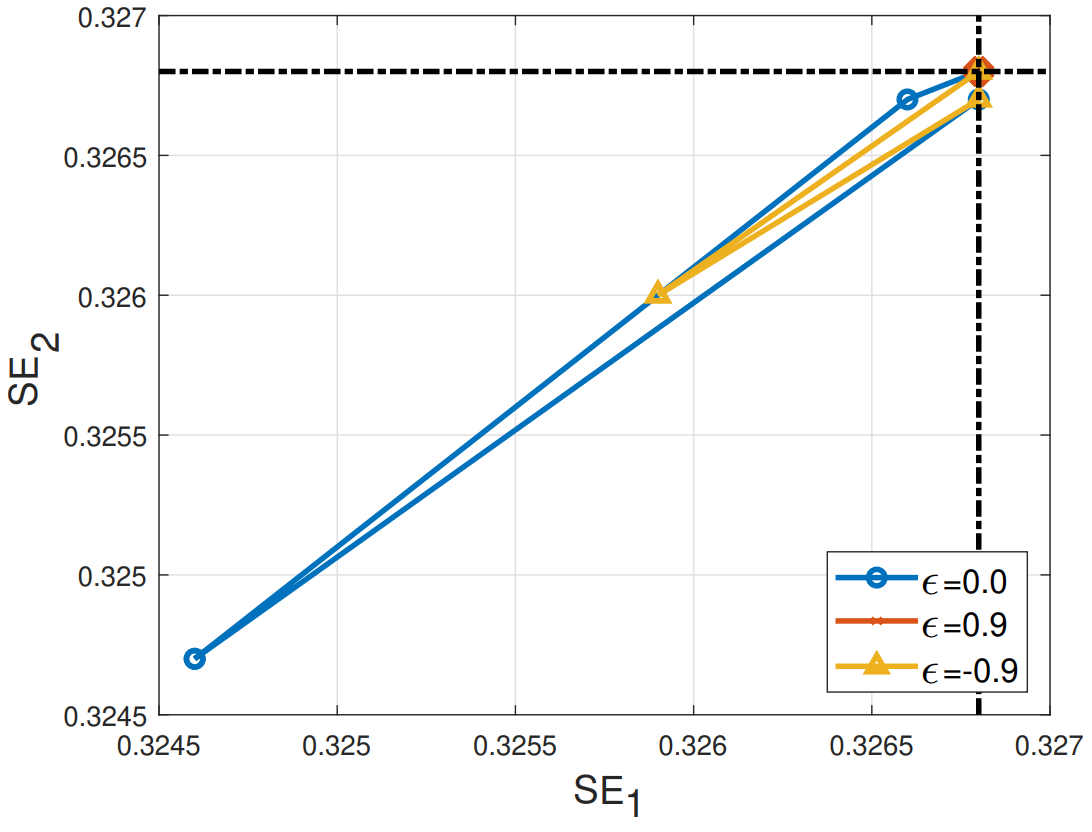} 
	\caption{The two user spectral efficiency for a rate $r=50/153$ DRF code over Guassian multicast channels with feedback, ($\eta_1=\eta_2=\infty$).} 
	\label{fig:AWGNBCSE}
\end{figure}

For better interpretation, we present the corresponding spectral efficiency values in Fig. \ref{fig:AWGNBCSE}. The spectral efficiency for receiver $r$ is calculated as
\begin{align}
    \mathrm{SE}_r=\frac{K \times (1-\mathrm{BLER}_r)}{L},
\end{align}
where $K$, and $L$ represent the number of transmitted bits, and the corresponding number of channel uses, respectively, whereas $\mathrm{BLER}_r$ is the block error rate at the output of decoder $r$ ($r=1,2$). The dotted black lines in Fig. \ref{fig:AWGNBCSE} represent the asymptotic spectral efficiency for each receiver, i.e. $50/153=0.3268$. This figure shows that, when the two noise variables $n^1$ and $n^2$ are positively correlated, we can achieve performance very close to the error-free spectral efficiency for both users. We lose some performance when the two noises are negatively correlated and the biggest performance loss occurs when they are independent. As expected, encoding becomes the most challenging when the noise variables are independent. Note that the DRF code can be similarly generalized to more than 2 receivers.


\section{Conclusions}\label{sec:Conclusions}
In this paper, we proposed a DNN-based error correction code for fading channels with output feedback, called deep SNR-robust feedback (DRF) code. The proposed encoder transmits the message along with a sequence of parity symbols, which are generated by an LSTM architecture based on the message as well as the observations of the past forward channel outputs available to the encoder with some additional noise. The decoder is implemented as a two-layer bi-directional LSTM architecture complemented with a SNR-aware attention mechanism. It is shown that the DRF code significantly improves over the previously proposed DNN-based codes in terms of the error rate as well as robustness to varying SNR values for AWGN channels with noisy feedback. Over fading channels, we showed that DRF codes can learn to efficiently use the knowledge of the instantaneous channel fading (available to the encoder through feedback) to reduce the overhead and complexity associated with channel estimation at the receiver. Finally, we generalized DRF codes to multicast channels with feedback, in which linear feedback codes are known to fall short of achieving the capacity. We showed that DRF codes can improve the reliability of both receivers simultaneously. DRF codes can be extended to many other types of channels, e.g., interference channels or relay channels with feedback, which we leave for future research.

\section*{Acknowledgement}
The authors would like to thank Dr. Yulin Shao from Imperial College London, Dr. Majid Nasiri Khormuji and Dr. Renaud-Alexandre Pitaval from Huawei Technologies, Sweden for insightful discussions and constructive comments on this manuscript.


\bibliographystyle{IEEEtran}
\bibliography{IEEEabrv,refs}

\end{document}